\documentclass[10pt,a4paper]{article}
\usepackage{epsfig}

\def\figsubcap#1{\par\noindent\centering\footnotesize(#1)}
\title{Scalar field phase dynamics in preheating}
\author{T. Charters${}^{\dag}$, A. Nunes${}^{\ddag}$, J. P. Mimoso${}^{\S}$}

\begin{document}
\maketitle

\begin{center}
{${}^{\dag}{}$ Departamento de  Mec\^anica/Sec\c c\~ao de Matem\'atica\\
Instituto  Superior  de  Engenharia  de  Lisboa\\  Rua  Conselheiro  Em\'{\i}dio
Navarro, 1,  P-1949-014 Lisbon, Portugal\\  Centro de F\'{\i}sica  Te\'{o}rica e
Computacional  da Universidade  de Lisboa  \\  Avenida Professor  Gama Pinto  2,
P-1649-003 Lisbon,  Portugal\\${}^{\S}$${}^{\ddag}$ Departamento de F\'{\i}sica,
Faculdade  de Ci\^encias  da Universidade  de  Lisboa \\  Centro de  F\'{\i}sica
Te\'{o}rica e Computacional da Universidade  de Lisboa \\ Avenida Professor Gama
Pinto  2,  P-1649-003   Lisbon,  Portugal  \\  ${}^\dag$tca@cii.fc.ul.pt,
${}^\ddag$anunes@ptmat.fc.ul.pt, ${}^\S$jpmimoso@cii.fc.ul.pt}
  
\end{center}

\begin{abstract} We study  the model of a massive  inflaton field $\phi$ coupled
to another scalar  filed $\chi$ with interaction term  $g^2\phi^2\chi^2$ for the
first stage of preheating.  We obtain the the behavior of the  phase in terms of
the iteration of a simple family  of circle maps. When expansion of the universe
is taken  into account the qualitative  behavior of the phase  and growth number
evolution is reminiscent of the behavior found in the case without expansion.
\end{abstract}

\section{Introduction}  The  reheating  mechanism  was  proposed  as  a  period,
immediately after  inflation, during which the inflaton  field $\phi$ oscillates
coherently  about  its  ground  state  and swiftly  transfers  its  energy  into
ultra-relativistic matter  and radiation, here modelled by  another scalar field
$\chi$.  Consider  the potential  $V(\phi)=1/2 m^2_\phi \phi^2$  and interaction
potential\cite{Kofman:1997b,Dolgov:1982th,Abbott:1982hn,Traschen:1990sw}
$V_{int}(\phi,\chi) = g^2 \phi^2\chi^2$. The  evolution of the flat FRW universe
is given by
\begin{eqnarray}
   \label{hubble} 3 H^2 =  \frac{8 \pi}{m_{pl}^2}\left( \frac{1}{2}\dot \phi^2 +
V(\phi)+\frac{1}{2}\dot \chi^2 + g^2\phi^2\chi^2\right),
\end{eqnarray}  where  $H=\dot R/R$  and  $R$ is  the  FRW  scalar factor.   The
equations of  motion in  a FRW  universe for a  homogeneous scalar  field $\phi$
coupled to the $k$-mode of the $\chi$ field are given by
\begin{eqnarray}
  \label{eqphigen}  \ddot\phi  +   3H\dot\phi  +  \left(m_\phi^2  +  g^2\chi_k^2
\right)\phi &=& 0,\\
  \label{eqchigen} \ddot\chi_k + 3H\dot\chi_k + \omega^2_k(t)\chi_k &=& 0,
\end{eqnarray}  where $\omega_k^2(t)  =  k^2  /R^2 +  g^2  \phi^2$.  The  number
density $n_k(t)$ of  particles with momentum $k$ can be  evaluated as the energy
of   that   mode  divided   by   the  energy   of   each   particle  $n_k(t)   =
\left(\vert\dot\chi_k\vert^2      +      \omega^2(t)\vert\chi_k\vert^2\right)/(2
\omega_k(t))-1/2$.

In this communication  we extend the formalism of  \cite{Kofman:1997b} to give a
full description  of the  dynamics of the  phase of  the field modes  $\chi _k$,
which in  turn determines the  evolution of the  growth factor $\mu _k$.   For a
complete reference see Charters \emph{et al.}\cite{Charters:2005}.

\section{Phase dynamics in Minkowski space-time}
\label{with.out.expansion}
\begin{figure}[t]%
\begin{center}
  \parbox{2.1in}{\epsfig{figure=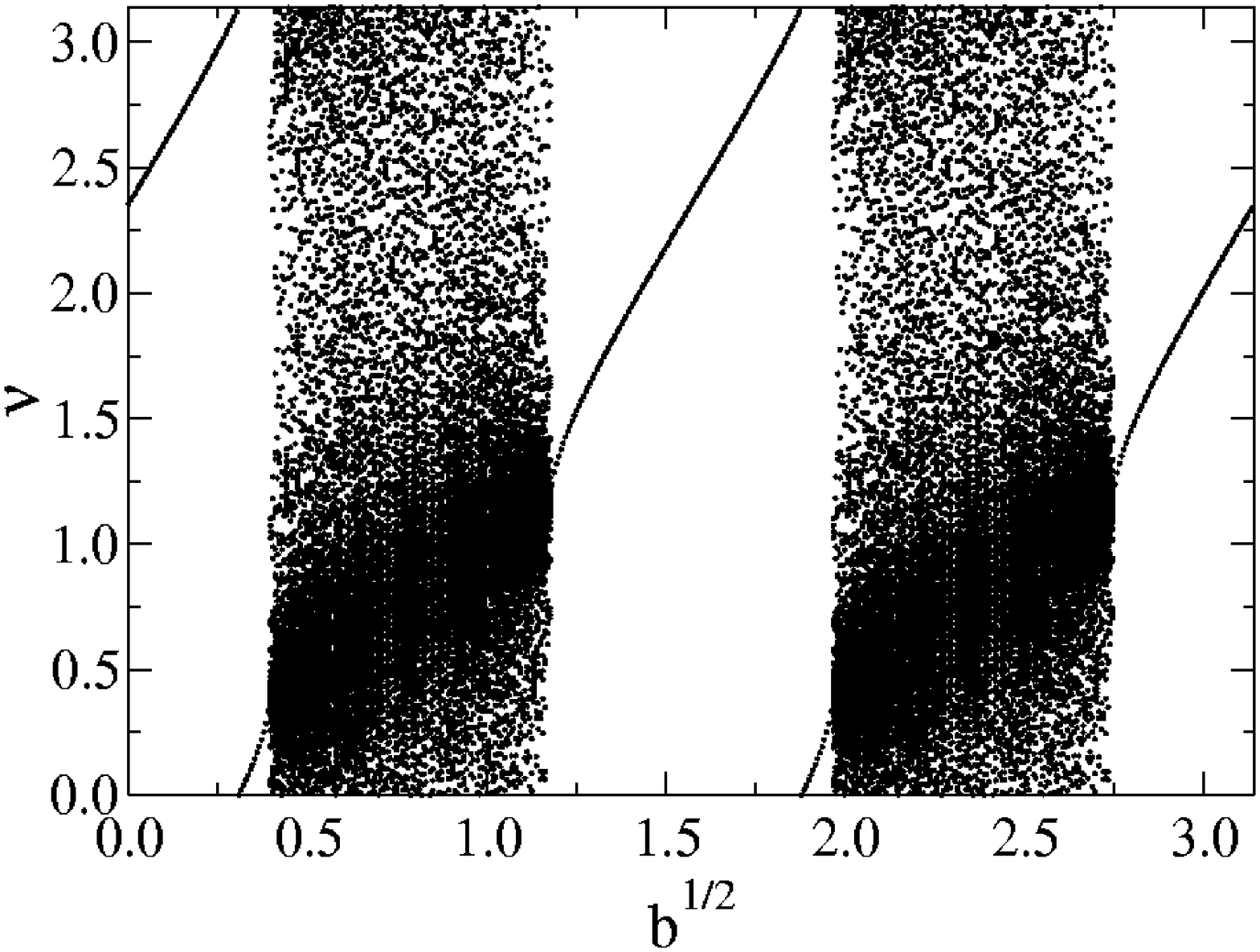,width=2.0in}\figsubcap{a}}
\hspace*{4pt}
  \parbox{2.1in}{\epsfig{figure=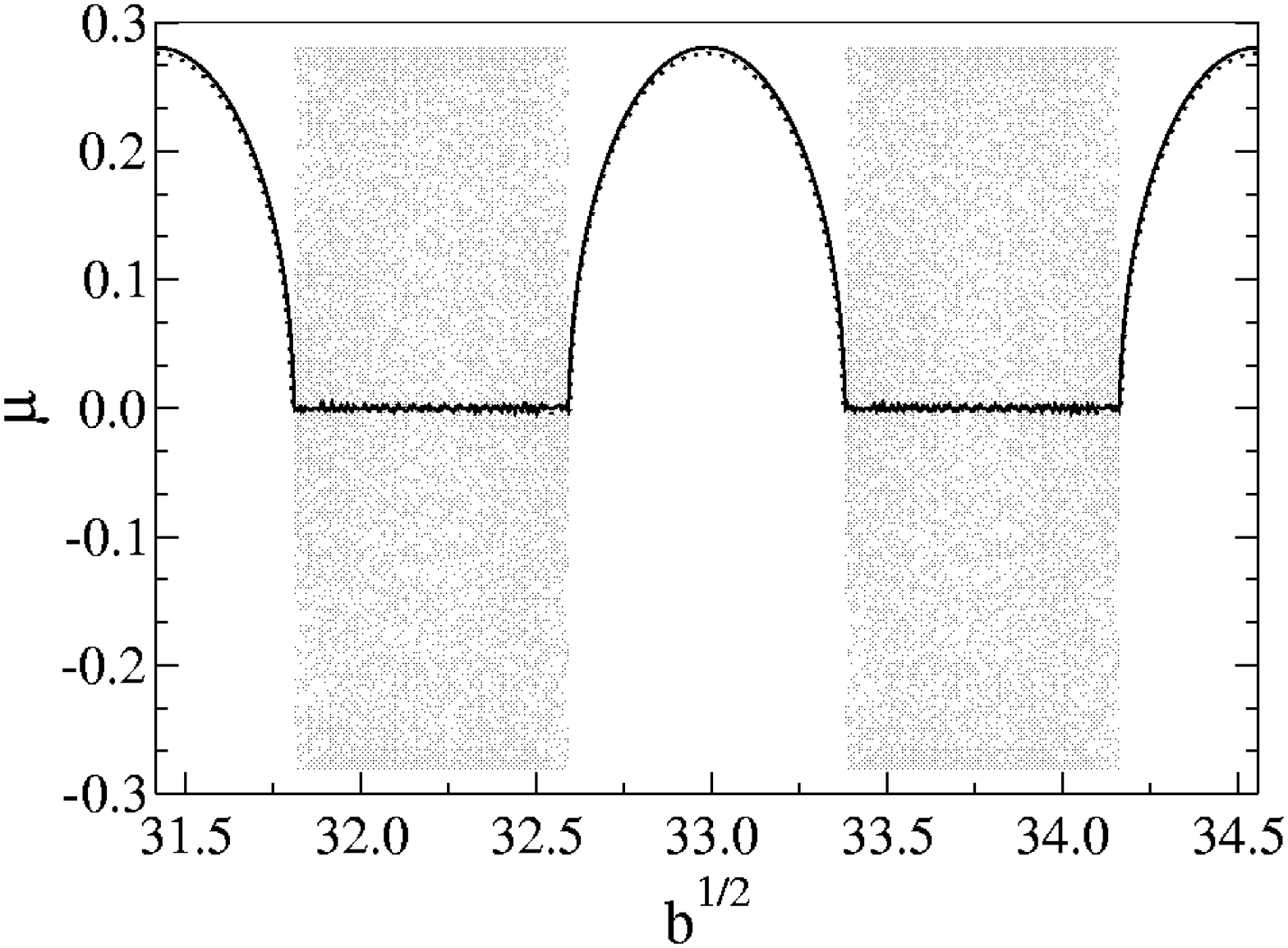,width=2.0in}\figsubcap{b}}
  \caption{(a)   Bifurcation   diagram   of    the   family   of   circle   maps
(\ref{phaseiter_k=0}) for $\sqrt{b}  \in [0, \pi]$. (b) The  asymptotic value of
$\mu  _0$  as  a function  of  $b$  for  $\sqrt{b}\in [10  \pi,11\pi]$  computed
analytically (full line) and numerically (dotted line). Also shown (in grey) are
all the values of $\mu_0^j$, $j=100,101,\ldots,200$.}
  \label{fig2}
\end{center}
\end{figure}

Typical  values  of  the  parameters  \cite{Kofman:1997b,Linde:90}  are  $g^2\le
10^{-6}$, $m=10^{-6}m_{pl}$, $A=\alpha m_{pl}$, where $0 < \alpha < 1$, and thus
$b\le \alpha ^2\times  10^{6}$.  In the present case we  have, $\sqrt{b} \gg 1$,
and it is possible to construct an approximate global solution
\begin{eqnarray}
\label{chi_j}
\chi_k^j(t;\alpha_k^j,\beta_k^j)=\frac{\alpha_k^j}{\sqrt{2\omega_k(t)}}e^{-     i
\int_0^t\omega_k(s)      d     s}+\frac{\beta_k^j}{\sqrt{2\omega_k(t)}}e^{     i
\int_0^t\omega_k(s) d s},
\end{eqnarray}  which is  valid except  in the  neighbourhood of  $t_j=  j \pi$,
$j=0,1,\ldots$.     $\alpha_k^j$    and    $\beta_k^j$   are    constants    and
$\omega_k^2(t)=a_k     +     b\sin^2(t)$     with     $a_k=k^2/m_\phi^2$     and
$b=g^2A^2/m_\phi^2$, and $A$ is the  constant amplitude of the field $\phi$, and
the    time   variable    is   now    $t\to   m_\phi    t$.     The   parameters
$(\alpha_k^j,\beta_k^j)$ for consecutive $j$  are determined by the behaviour of
the solution of (\ref{eqchigen}) in Minkowski space-time for $t$ close to $t_j$.
In terms  of the phase  $\nu _k^j  = \arg \beta_k^j  + \theta_k^j$ of  the field
$\chi _k$ when $t=t_j$, one gets \cite{Kofman:1997b}
\begin{eqnarray}
\label{phaseiter}   \nu_k^{j+1}   =   \theta    (b,   \kappa)   +   \arg   \left
(\sqrt{1+\rho_\kappa^2} e^{-  i \varphi_\kappa} e^{  i \nu_k^j} -  i \rho_\kappa
e^{- i \nu_k^j} \right ),
\end{eqnarray}   where  $\theta   (b,  \kappa   )=\int_0^{\pi}\omega(s)   d  s$,
$\rho_{\kappa}  =   \exp(-\pi  \kappa^2/2)$,  $\kappa  ^2   =  a_k/\sqrt{b}$,  $
\varphi_{\kappa}       =\arg(\Gamma((1+      i       \kappa       ^2)/2))      +
\kappa^2/2(1+\ln(2/\kappa^2))$, and  $\kappa =k  / \sqrt{A g  m_{\phi}}$.  Since
$n_k=\vert\beta_k\vert^2$,   the  growth   index   $\mu_\kappa^j$,  defined   by
$n_k^{j+1}=n_k^j\exp(2 \pi  \mu_\kappa^j)$, is given  by \cite{Kofman:1997b}, in
terms of the phase  $\nu _k^j = \arg \beta_k^j + \theta_k^j$  of the field $\chi
_k$ when $t=t_j$,
\begin{eqnarray}
\label{munu.phase} \mu_\kappa^j  = \frac{1}{2 \pi}\ln\left(1  + 2\rho_\kappa^2 -
2\rho_\kappa\sqrt{1+\rho_\kappa^2}\sin(-\varphi_\kappa +2\nu_k^j ) \right).
\end{eqnarray} The properties of the family (\ref{phaseiter}) is best understood
by looking at the behavior (see Figure \ref{fig2}) of the family $P_{b,0}(\nu )$
parametrised by $\sqrt{b}$\cite{Charters:2005},
\begin{eqnarray}
\label{phaseiter_k=0} P_{b,  0}(\nu) =  2 \sqrt{b} +  \arctan\frac{\sqrt{2} \sin
\nu - \cos \nu}{\sqrt{2} \cos \nu - \sin \nu}.
\end{eqnarray} The map has a  two dynamical regimes, a strongly attractive fixed
point, and random oscillations around a mean value (see Figure \ref{fig2}).

\section{Dynamics in an expanding universe}
\label{with.expansion}

\begin{figure}[t]%
\begin{center}
  \parbox{2.1in}{\epsfig{figure=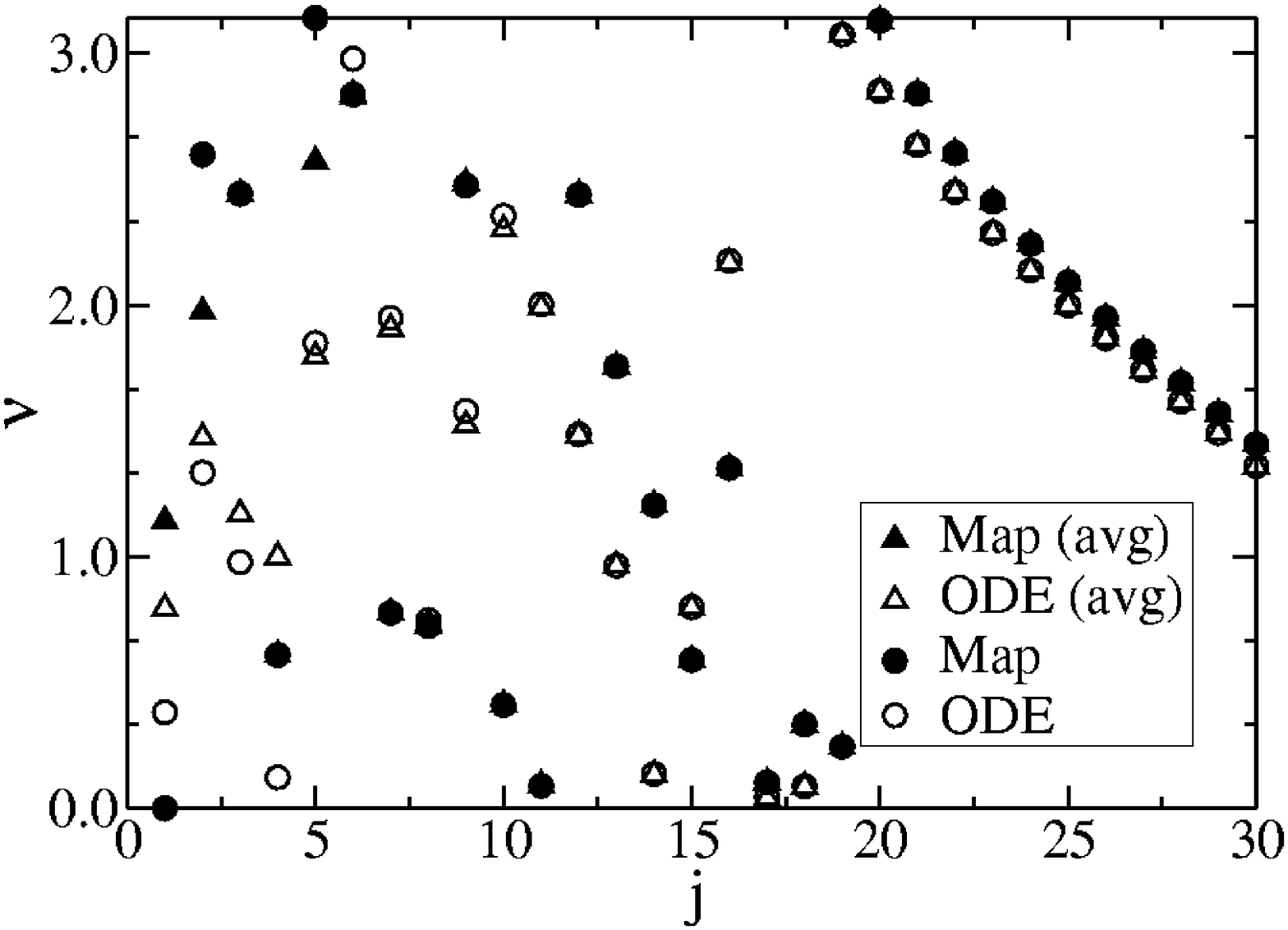,width=2.0in}\figsubcap{a}}
\hspace*{4pt}
  \parbox{2.1in}{\epsfig{figure=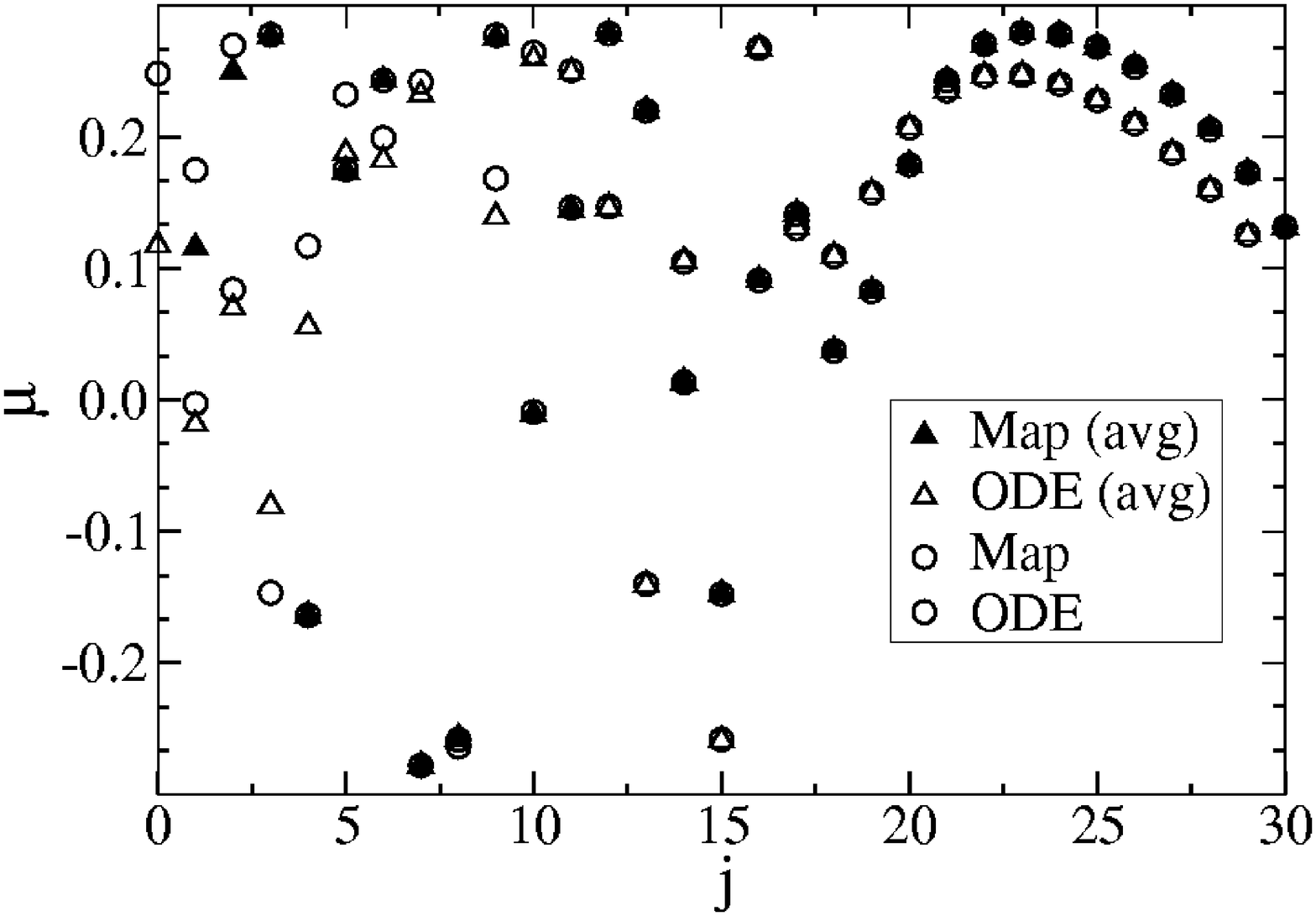,width=2.0in}\figsubcap{b}}
  \caption{For  $b_0=  5  \times  10^3$  and  $\kappa_0=0.1$  and  with  initial
conditions  $n_k^0 =  1/2$ and  $\nu _k^0  =  0$ (a)  The phase  (b) The  growth
factor. For the  same initial conditions and parameter  values, the iteration of
equations  are  plotted  as  full  circles, and  the  numerical  integration  of
equations  and  are  plotted  as  open  circles. This  were  carried  out  until
$\sqrt{b(t_j)} \approx 1$.}
  \label{fig3}
\end{center}
\end{figure}

In the  first stage of preheating,  that ends when $n_\chi  (t) \approx m_\phi^2
\Phi(t)/g$,  where $\Phi(t)$  is the  varying  amplitude of  the inflaton  field
$\phi$,    equations     (\ref{hubble}),    (\ref{eqphigen})    decouple    from
(\ref{eqchigen}),  and the  evolution of  the inflaton  field and  of  the scale
factor  $R(t)$ is  given by  \cite{Kofman:1997b} $  \phi (t)=  \Phi(t)  \sin t$,
$\Phi(t)=m_{pl}/(3(\pi  /2   +  t))$,   $R(t)=  (2  t/\pi   )^{2/3}$.   Equation
(\ref{eqchigen}) can be reduced to the form of an harmonic oscillator the change
of  variable  $X_k  = R^{3/2}\chi_k$,  with  frequency  given  by $  \varpi^2  =
k^2/(m_\phi^2R(t)^2)  +  g^2\phi(t)^2/m_\phi^2   +  \delta/m_\phi^2$  and  where
$\delta/m_\phi^2 \ll 1$.
 The  preheating period  ends  when  $g \Phi(t)/m_\phi  \simeq  1$, and,  during
preheating, the  rate of variation of  those parameters and  the oscillations of
the inflaton field are much slower than the oscillations of the $\chi _k$ modes.
The changes in occupation numbers $n_k$ will occur at $t=j \pi$ with exponential
growth rate  given by  (\ref{munu.phase}), provided we  take $\kappa_j=k/(R(t_j)
\sqrt{g m_\phi \Phi(t_j)})$  and $\sqrt{b_j} = g \Phi  (t_j)/m_\phi$ (see Figure
(\ref{fig3})).

\section{Conclusions}
 We use the
theory of developed by Kofman \emph{et al.}\cite{Kofman:1997b} to 
 show that the  phase dynamics of the modes $\chi _k$ in Minkowski space-time is given by the
properties of a simple family of circle maps.
 In the  case of  an expanding  universe the
qualitative behaviour of the phase and growth number evolution is reminiscent of
the behavior found in the case without expansion.


\end{document}